\documentclass[twocolumn,prl,amsmath,amssymb,showpacs]{revtex4}
\usepackage{graphicx}
\usepackage{dcolumn}
\usepackage{bm}
\usepackage{latexsym}
\usepackage{amstext}
\usepackage{amsxtra}

\newcommand{\bs}{\boldsymbol}
\newcommand{\op}{\omega_\perp}
\newcommand{\Oef}{\Omega_{\rm eff}}
\newcommand{\Oap}{\Omega_{\rm stir}}
\newcommand{\OapI}{\Omega_{\rm stir}^{(1)}}
\newcommand{\OapII}{\Omega_{\rm stir}^{(2)}}
\newcommand{\rp}{r_{\perp}}

\begin{document}
\title{Fast rotation of an ultra-cold Bose gas}
\author{Vincent Bretin, Sabine Stock, Yannick Seurin,
and Jean Dalibard} \affiliation{Laboratoire Kastler Brossel$^{*}$,
24 rue Lhomond, 75005 Paris, France}
\date{July 18, 2003}

\begin{abstract}
We study the rotation of a $^{87}$Rb Bose-Einstein condensate
confined in a magnetic trap to which a focused, off resonant
gaussian laser beam is superimposed. The confining potential is
well approximated by the sum of a quadratic and a quartic term
which allows to increase the rotation frequency of the gas above
the trap frequency. In this fast rotation regime we observe a
dramatic change in the appearance of the quantum gas. The vortices
which were easily detectable for a slower rotation become much
less visible, and their surface density is well below the value
expected for this rotation frequency domain. We discuss some
possible tracks to account for this effect.
\end{abstract}


\maketitle

The fast rotation of a  macroscopic quantum object often involves
a dramatic change of its properties. For rotating liquid $^4$He,
superfluidity is expected to disappear for large rotation
frequencies \cite{Donnelly91}. A similar effect occurs for type-II
superconductors placed in a magnetic field, which loose their
superconductivity when the field exceeds a critical value
\cite{Tinkham}. For the last few years it has been possible to set
gaseous Bose-Einstein condensates into rotation, by nucleating in
the gas quantized vortices which arrange themselves in a
triangular lattice
\cite{Matthews99,Madison00,Ketterle1,Hodby01,Haljan01}. The
theoretical investigation of the properties of these fast rotating
gases has led to several possible scenarios. Depending on the
confinement of the gas, its dimensionality and the strength of the
interactions, theoretical predictions involve the nucleation of
multiply charged vortices
\cite{Fetter01,Lundh02,Kasamatsu02,Kavoulakis03}, melting of the
vortex lattice \cite{Sinova02,Fisher03}, or effects closely
connected to Quantum Hall physics
\cite{Cooper01,Paredes01,Ho01,Regnault03}.

The confinement of cold atomic gases is generally provided by a
magnetic trap which creates an isotropic transverse potential
$m\op^2\rp^2/2$ in the plane perpendicular to the rotation axis
$z$ (we set $x^2+y^2=\rp^2$). As the rotation frequency $\Omega$
varies from $0$ to $\op$, the number of quantized vortices
increases \cite{Raman01}. The limiting case of $\Omega=\op$, e.g.
fast rotation, is singular: the confinement vanishes since the
trapping force is compensated by the centrifugal force. In
classical terms, the only remaining force on the particle is the
Coriolis force. In quantum terms this corresponds to a gauge field
$\bs A(\bs r)=m\bs \Omega \times \bs r/2$. This is analog to the
physics of charged particles in a uniform magnetic field and one
expects to recover for weak interactions an energy spectrum with
Landau levels separated by $2\,\hbar\omega_\bot$. However the
absence of confinement in a pure harmonic trap rotating at
$\Omega=\op$ makes this study experimentally delicate
\cite{Rosenbusch01}. The Boulder group recently reached
$\Omega=0.99\;\op$ using evaporative spin-up \cite{Boulder03}.

In the present paper, following suggestions in
\cite{Fetter01,Lundh02,Kasamatsu02,Kavoulakis03} we study the
rotation of a Bose-Einstein condensate in a trap whose potential
is well approximated by a superposition of a quadratic and a small
quartic potential:
 \begin{equation}
V(\bs r)\simeq \frac{1}{2}m\omega_z^2 z^2+\frac{1}{2}m\op^2 \rp^2
+\frac{1}{4}k\rp^4 \qquad (k>0)\ .
 \label{potential}
 \end{equation}
We can thus explore with no restriction the domain of rotation
frequencies $\Omega$ around $\op$. Our results show a strong
dependance of the behavior of the gas with $\Omega$. For
$\Omega/\op<0.95$ we observe a regular vortex lattice. This
lattice gets disordered when $\Omega$ increases. For $\Omega>\op$
the number of detectable vortices is dramatically reduced although
we have clear evidences that the gas is still ultra-cold and in
fast rotation. We conclude this Letter by proposing some possible
explanations for this behavior.

Our $^{87}$Rb condensate is formed by radio-frequency evaporation
in a combined magnetic+laser trap. The pure magnetic trap provides
a harmonic confinement along the 3 directions with the frequencies
$\op^{(0)}/2\pi=75.5$~Hz and $\omega_z/2\pi=11.0$~Hz. We
superimpose a blue detuned laser (wavelength 532 nm) propagating
along the $z$ direction to provide the quartic term in the
confinement. The potential created by the laser is
 \begin{equation}
U(\rp)=U_0\,\exp\left(-\frac{2\rp^2}{w^2}\right)\simeq U_0
-\frac{2U_0}{w^2}\rp^2+\frac{2U_0}{w^4}\rp^4\;.
 \label{laser}
 \end{equation}
The laser's waist is $w=25\;\mu$m and its power is 1.2~mW. The
first term $U_0$ ($\sim k_B\times 90$~nK) in the right hand side
of (\ref{laser}) is a mere shift of the energy scale. The second
term is a correction of the transverse trapping frequency; we get
for the combined magnetic-laser trap $\op/2\pi=65.6$~Hz
\cite{anisotropy}.  The last term in (\ref{laser}) corresponds to
the desired quartic confinement, with
$k=2.6\,(3)\times10^{-11}$~J.m$^{-4}$. The expansion in
(\ref{laser}) is valid for $\rp \lesssim w/2$, which is indeed the
case for all data presented here.

\begin{figure*}[t]
 \includegraphics{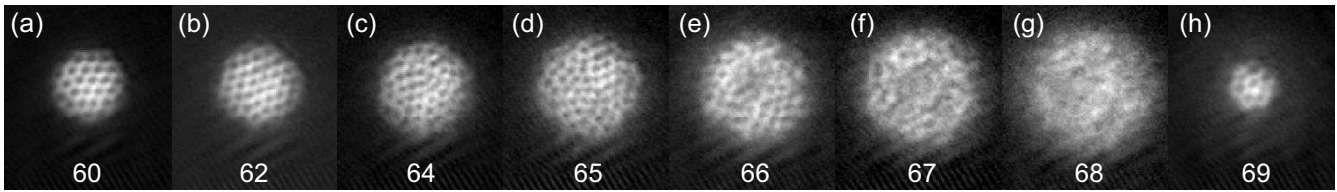}
 \caption{Pictures of the rotating gas taken along the
rotation axis after 18~ms time of flight. We indicate in each
picture the stirring frequency $\OapII$ during the second stirring
phase ($\op/2\pi=65.6$~Hz). The real vertical size of each image
is 306\;$\mu$m.}
 \label{fig:images}
\end{figure*}

We start the experimental sequence with a quasi-pure condensate
that we stir using an additional laser beam, also propagating
along $z$ \cite{Madison00}. This laser stirrer creates an
anisotropic potential in the $xy$ plane which rotates at a
frequency $\Oap$. To bring the condensate in rotation at a
frequency close to $\op$, we use two stirring phases. First we
choose $\OapI\simeq \op/\sqrt{2}$, so that the stirring laser
resonantly excites the transverse quadrupole mode $m=+2$ of the
condensate at rest \cite{Madison01}. The duration of this first
excitation is 300~ms and we then let the condensate relax for
400~ms in the axisymmetric trap. This procedure sets the
condensate in rotation, with a vortex lattice containing typically
15 vortices.

We then apply the laser stirrer for a second time period, now at a
rotation frequency $\OapII$ close to or even above $\op$. For the
condensate with 15 vortices prepared during the first phase, the
transverse quadrupole mode is shifted to a higher frequency and
can now be resonantly excited by a stirrer rotating at $\OapII\sim
\op$ \cite{Zambelli98,Haljan01}. The second stirring phase lasts
for 600~ms. It is followed by a 500~ms period during which we let
the condensate equilibrate again in the axisymmetric trap. At this
stage the number of trapped atoms is $3\times 10^5$.

We find that this double nucleation procedure is a reliable way to
produce large vortex arrays and to reach effective rotation
frequencies $\Oef$ around or above $\op$. In particular it is much
more efficient than the procedure which consists in applying
permanently the stirring laser while increasing continuously its
frequency. It has one drawback however: the effective rotation
frequency $\Oef$ of the gas after equilibration might differ
significantly from the rotation frequency $\OapII$ that we apply
during the second stirring phase. The role of the stirring phase
is to inject angular momentum in the system, and the rotation
frequency of the atom cloud may subsequently change if the atom
distribution (hence the moment of inertia) evolves during the
final equilibration phase.

The rotating atom cloud is then probed destructively by switching
off the confining potential, letting the cloud expand during
$\tau=18$~ms and performing absorption imaging. We take
simultaneously two images of the atom cloud after time of flight
(TOF). One imaging beam is parallel to the rotation axis $z$ and
the other one is perpendicular to $z$ \cite{Rosenbusch02}.

\begin{figure}
 \includegraphics{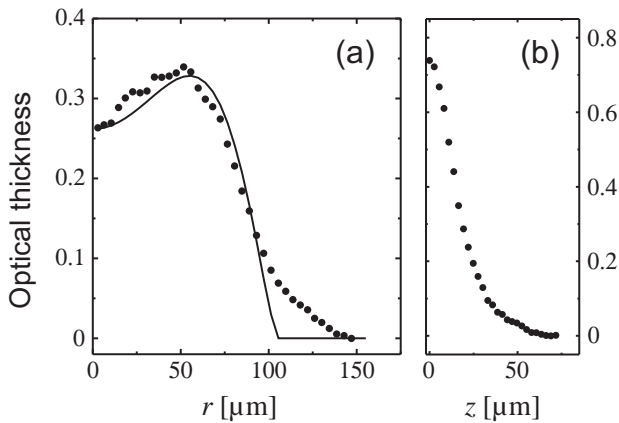}
 \caption{
Optical thickness of the atom cloud after time-of-flight for
$\OapII/2\pi=66$~Hz. (a) Radial distribution in the $xy$ plane of
Fig.~\ref{fig:images}e. Continuous line: fit using the
Thomas-Fermi distribution (\ref{TF}). (b) Distribution along the
$z$ axis averaged over $|x|<20\;\mu$m (imaging beam propagating
along $y$).}
 \label{fig:fit}
\end{figure}

Fig.~\ref{fig:images}a-h shows a series of images taken along the
$z$ axis and obtained for various $\OapII$ around $\op$.  When
$\OapII$ increases, the increasing centrifugal potential weakens
the transverse confinement. This leads to an increasing radius in
the $xy$ plane. As it can be seen from the transverse and
longitudinal density profiles of Fig.~\ref{fig:fit}, obtained for
$\OapII/2\pi=66$~Hz, the gas after TOF has the shape of a flat
pancake. The limit of our method for setting the gas in fast
rotation is shown on the image (h) at the extreme right of
Fig.~\ref{fig:images}. It corresponds to $\OapII/2\pi=69$~Hz, far
from the quadrupole resonance of the condensate after the first
nucleation. In this case we could not bring the gas in the desired
fast rotation regime.

For the pictures (e,f) obtained with $\OapII/2\pi=66$ and $67$~Hz,
the optical thickness of the cloud has a local minimum in the
center; this indicates that the confining potential in the
rotating frame, $V_{\rm rot}(\bs r)=V(\bs r)-m\Oef^2 \rp^2/2$, has
a Mexican hat shape. The effective rotation frequency $\Oef$ thus
exceeds $\op$. The striking feature of these images is the small
number of visible vortices which seems to conflict with a large
value of $\Oef$. The main goal of the remaining part of this
Letter is to provide more information on this puzzling regime.

In order to analyze quantitatively the pictures of
Fig.~\ref{fig:images}, we need to model the evolution of the cloud
during TOF. For a pure harmonic potential ($k=0$), a
generalization of the analysis of \cite{Castin96} to the case of a
rotating condensate shows that the expansion in the $xy$ plane is
well described by a scaling of the initial distribution by the
factor $(1+\op^2\tau^2)^{1/2}$ \cite{Chevy01}. We assume here that
this is still approximately the case for a condensate prepared in
a trap with a non zero quartic term $k\rp^4/4$.

\begin{figure}
 \includegraphics{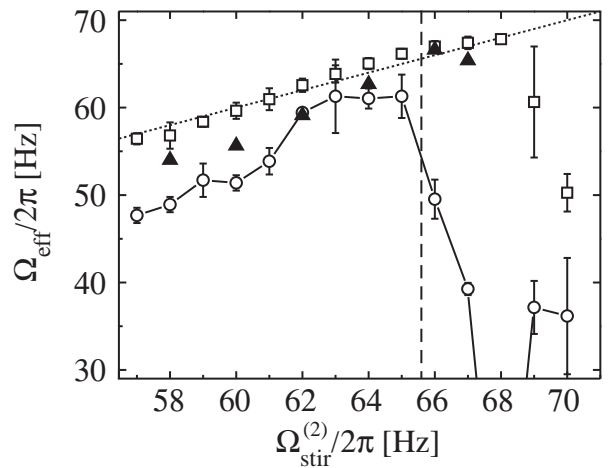}
\caption{Effective rotation frequency $\Oef$  as a function of the
stirring frequency $\OapII$. $\Box$: values deduced from the fit
using the Thomas-Fermi distribution (\ref{TF}). {\large $\circ$}:
values obtained by measuring the vortex density in the TOF
pictures. $\blacktriangle$: values obtained using surface wave
spectroscopy. The bars indicate standard deviations.}
 \label{fig:omega}
\end{figure}

We have analyzed a series of 60 images such as those of
Fig.~\ref{fig:images}, assuming an initial atomic distribution
given by the Thomas-Fermi law:
 \begin{equation}
n(\bs r)=\frac{m}{4\pi \hbar^2 a}\left(\mu- V_{\rm rot}(\bs r)
\right)\ ,
 \label{TF}
 \end{equation}
where $a=5.2$~nm is the scattering length characterizing the
atomic interactions and $\mu$ the chemical potential. The optical
thickness for the imaging beam propagating along $z$, proportional
to the column atomic density in the $xy$ plane, varies as
$(\alpha+\beta r_\bot^2+\gamma r_\bot^4)^{3/2}$, where
$\alpha,\beta,\gamma$ can be expressed in terms of the physical
parameters of the problem \cite{imagetrans}. An example of the
fit, which takes $\mu$ and $\Oef$ as adjustable parameters, is
given in Fig.~\ref{fig:fit}a. The agreement is correct, though not
as good as for condensates confined in purely harmonic traps. This
may be a consequence of the approximate character of the scaling
transform that we use to describe the TOF evolution. The resulting
values for $\Oef$ as a function of $\OapII$ are given in
Fig.~\ref{fig:omega}. We find $\Oef \simeq \OapII$ for stirring
frequencies $\leq 68$~Hz. Note that for $\OapII/2\pi=68$~Hz (image
Fig.~\ref{fig:images}g) the quality of the fit is comparatively
poor due to local inhomogeneities of the atom cloud.

 From the value of $\mu$ given by the fit, we recover the atom
number \cite{Fetter01}. For the largest measured rotation
frequency, the Thomas-Fermi distribution (\ref{TF}) corresponds to
a nearly spherical atom cloud before TOF (diameter 29~$\mu$m in
the $xy$ plane and length 34~$\mu$m along $z$).

A second determination of the effective rotation frequency $\Oef$
of the condensate is provided by the vortex surface density after
TOF. Assuming that the vortex pattern is scaled by the same factor
as the condensate density, we deduce the vortex density $\rho_{v}$
before TOF, hence the rotation frequency $\Oef=\pi\hbar\rho_v/m$
\cite{Raman01}. For $\OapII<\op$, the value of $\Oef$ deduced in
this way and plotted in Fig.~\ref{fig:omega} is in fair agreement
with the one deduced from the fit of the images. On the contrary,
for $\OapII/2\pi=66-68$~Hz the number of distinguishable vortices
is much too low to account for the rotation frequencies determined
from the fits of the TOF images.

In order to gather more information on the rotational properties
of the gas, we now study the two transverse quadrupole modes
$m=\pm 2$ of the gas. We recall that these two modes have the same
frequency for a non rotating gas, due to symmetry. For a rotating
gas in the hydrodynamic regime, the frequency difference
$\omega_{+2}-\omega_{-2}$ is proportional to the ratio between the
average angular momentum of the gas and its moment of inertia,
which is nothing but the desired effective rotation frequency
$\Oef$ \cite{Zambelli98}. The strength of this approach is that it
does not make any assumption on the expansion during TOF.

To study these transverse quadrupole modes, we briefly illuminate
the rotating gas using the laser stirrer, now with fixed axes
\cite{Chevy00}. This excites a superposition of the modes $m=\pm
2$ with equal amplitudes. We then let the cloud evolve freely in
the trap for an adjustable duration and we perform the TOF
analysis. From the time variation of both the ellipticity of the
cloud in the $xy$ plane and the inclination of its eigenaxes, we
deduce $\omega_{\pm 2}$, hence the effective rotation frequency
$\Oef=(\omega_{+2}-\omega_{-2})/2$. The corresponding results are
plotted in Fig.~\ref{fig:omega}. They are in good agreement with
the results obtained from the fits of the images: $\Oef \simeq
\OapII$. Consequently they conflict with the rotation frequency
that one derives from the vortex surface density when
$\OapII/2\pi=66-68$~Hz.

To explain the absence of visible vortices in the regime
$\OapII/2\pi=66-68$~Hz, one could argue that the gas might be
relatively hot, hence described by classical physics. The
equilibrium state should then correspond to rigid body rotation.
However all experiments shown here are performed in presence of
radio-frequency (rf) evaporative cooling. The rf is set 24~kHz
above the value which empties the trap and it eliminates all atoms
crossing the horizontal plane located at a distance $x_{\rm
ev}\sim 19\;\mu$m below the center (one-dimensional evaporation).
>From evaporative cooling theory \cite{Ketterle96,Luiten96} we know
that the equilibrium temperature $T$ of the rotating gas is a
fraction of the evaporation threshold $V_{\rm rot}(x_{\rm ev})\sim
32$~nK for $\Oef/2\pi=67$~Hz \cite{gaussian}. This indicates that
$T$ is in the range of $5-15$~nK, well below the critical
temperature at this density (180~nK for an estimated density
$3\times10^{13}$~cm$^{-3}$) \cite{evaporation}. This low
temperature is also confirmed by the small decay rates of the
quadrupole modes ($\sim 20$~s$^{-1}$), characteristic of $T \ll
T_c$.

The estimated temperature $T$ is of the order of the splitting
between Landau levels $2\hbar\omega_\bot/k_B=6.3$~nK, so that only
the first two or three levels are appreciably populated. For each
Landau level, the rf evaporation eliminates states with an angular
momentum $L_z/\hbar > (x_{\rm ev}/a_{\rm ho})^2\sim 200$, where
$a_{\rm ho}=\sqrt{\hbar/(m\omega_\bot)}$ is the ground state's
size of the harmonic oscillator with frequency $\op$.

A first possibility to interpret our data consists in assuming
that the whole gas shown in Fig.~\ref{fig:images}e-g is rotating,
that it is in the degenerate regime, and that it cannot be
described by a single macroscopic wave function since the number
of distinguishable vortices is too low to account for the measured
rotation frequency. In this point of view, the state of the system
is therefore more complex than a single Hartree state. A
theoretical analysis along this line, involving the formation of
composite bosons, has been very recently proposed in
\cite{Akkermans03}.

A second possible explanation is that the vortices are still
present for $\Omega\geq \op$, but that the vortex lines are
strongly tilted or bent, so that they do not appear as clear
density dips in the images of Fig.~\ref{fig:images}e-g. This could
be for example a consequence of the residual anisotropy of the
trapping potential in the $xy$ plane \cite{anisotropy}. In an
experiment performed at $\Omega=0.95\,\op$, the Boulder group has
shown that a static anisotropy of $\sim 4$\% reduces considerably
the vortex visibility over a time scale of $0.8$~s
\cite{Engels02}. However the residual anisotropy of our trap is
$<1$\% and this effect should be limited, unless the vortex
lattice becomes notably more fragile as $\Omega$ approaches $\op$.
The distortion of the vortex lattice could also result from a
strong increase of the cristallization time for $\Omega \sim \op$
which would then exceed the 500~ms equilibration time
$\tau_\mathrm{eq}$. To check this effect, we have increased
$\tau_\mathrm{eq}$ up to 2~s without noticing any qualitative
change in the atom distribution (for larger $\tau_\mathrm{eq}$ the
gas slows down significantly).

A third possibility is that a small fraction of the gas has
stopped rotating during the standard 500~ms equilibration time,
and can therefore be at a higher temperature \cite{evaporation}.
This small fraction could decrease the visibility of the vortices
present in the condensed part of the cloud. The existence of such
a fraction could be tested by a quantitative analysis of the
density profiles of Fig.~\ref{fig:fit}ab. Such an analysis should
take into account the TOF evolution of an interacting atomic cloud
initially confined in a non harmonic potential.

To summarize we have presented in this Letter a direct evidence
for a qualitative change in the nature of a degenerate Bose gas
when it is rotated around and above the trapping frequency. We
plan to complement this study by an analysis of other modes of the
rotating gas, such as the transverse breathing mode
\cite{Stock03}. Several other interesting phenomena have been
predicted and remain to be investigated experimentally for this
type of potential: existence of several phases involving either a
vortex array with a hole, or a giant vortex
\cite{Fetter01,Lundh02,Kasamatsu02,Fisher03,Kavoulakis03}. Another
extension of the present work consists in transposing the
experimental scheme to a 2D geometry, where the motion along $z$
would be frozen. The situation would then be the bosonic analog of
the situation leading to the quantum Hall effect, provided the
effective rotation frequency of the gas is precisely adjusted to
the trap frequency.

{\acknowledgments We thank P. Rosenbusch for participation in
earlier stages of this experiment, and S. Stringari, L.
Pitaevskii, G. Baym and the ENS group for useful discussions. This
work is partially supported by CNRS, Coll\`{e}ge de France,
R\'{e}gion Ile de France, DAAD, DGA, DRED and EU (CQG network
HPRN-CT-2000-00125).}


\begin{thebibliography}{99}
\bibitem{anisotropy}
We measure an oscillation frequency of 65.4\;(3)~Hz along the
vertical axis $x$ and 65.9\;(3)~Hz along the horizontal axis $y$.
\bibitem{imagetrans}
Such a simple formula could not be obtained for the column density
along $y$, hence the absence of a fitting function in
Fig.~\ref{fig:fit}b.
\bibitem{gaussian}
For a precise evaluation of the evaporation threshold, one must
keep the gaussian variation of $U(r_\bot)$ in Eq.~(\ref{laser})
rather than using its quadratic+quartic approximation.
\bibitem{evaporation}
The rotation of the gas is crucial for establishing a link between
the evaporation radius and the temperature. For a gas at rest, the
evaporation threshold is $V(x_{\rm ev})$ instead of $V_{\rm
rot}(x_{\rm ev})$, which corresponds to a much higher temperature
(100-200~nK).
\end{thebibliography}
\end{document}